\documentstyle[preprint,aps,eqsecnum,epsfig]{revtex}
\tightenlines
\def\pmb#1{\setbox0=\hbox{#1}%
     \kern-.025em\copy0\kern-\wd0
      \kern.05em\copy0\kern-\wd0
       \kern-.025em\raise.0433em\box0}

\def\beq{\begin{equation}}
\def\eeq{\end{equation}}
\def\bea{\begin{eqnarray}}
\def\eea{\end{eqnarray}}
\begin{document}
\title{Kinetic-theory description of isoscalar dipole modes}
\author{ V. I. Abrosimov$^{\rm a}$, A.Dellafiore $^{\rm b}$ and
F.Matera$^{\rm b}$}
\address{  $^{\rm a}$ \it Institute for Nuclear Research, 03028 Kiev,
Ukraine\\
 $^{\rm b}$ \it Istituto Nazionale di Fisica Nucleare and Dipartimento
di Fisica,\\
Universita' di Firenze, Largo E.Fermi 2, ~50125 Firenze, Italy}
%
\maketitle

\begin{abstract}
A semiclassical model, based on a solution of the Vlasov equation for
finite systems with moving-surface, is employed to
study the isoscalar dipole modes in nuclei. It is shown that, by taking
into account the surface degree of freedom, it is possible to obtain an
exact treatment of the centre of mass motion. It is  also shown that a
method often used to subtract the spurious strength in RPA
calculations does not always give the correct result.  An
alternative analytical formula for the intrinsic strength function is derived
in a simple confined-Fermi-gas model. In this model the intrinsic isoscalar
dipole strength displays essentially a two-resonance structure, hence
there are two relevant modes. The interaction between nucleons
couples these two modes and changes the compressibility of the
system, that we determine by fitting the monopole resonance energy. The
evolution of the dipole strength profile is
studied as a function of the compressibility of the nuclear fluid.
Comparison with recent dipole data shows some qualitative agreement
for a soft equation of state,
but our model fails to reconcile the monopole and dipole data.
\end{abstract}
\vspace{.5 cm}
\vspace{.5 cm}

\section{Introduction}

The isoscalar dipole mode, excited by the operator
\beq
\label{spur}
O=\sum_{i}^{A} r_{i}^{3}Y_{1M}(\hat{\bf r}_{i})\,,
\eeq
is particularly interesting because it can give information on the
compressibility of nuclear matter \cite{har,str,gar} (Ref.\cite{gar}
contains a review of experimental and theoretical studies of
the isoscalar dipole resonance). This information is
complementary to that given by the monopole ``breathing'' mode
\cite{bla,ber}.

>From the theoretical point of view the study of the isoscalar dipole
mode is complicated by the fact that the operator
(\ref{spur}) can excite also the centre of mass (c.m.) motion.
If the model employed lacks translation invariance, the c.m. excitation may
appear as spurious strength in the calculated response.
It is well known that in a perfectly self-consistent random-phase
approximation (RPA) the problem of spurious strength does
not arise since the c.m. strength is concentrated at zero energy,
however, in practice, even a small lack of self-consistency can give
substantial spurious
dissipation at positive energy (see e.g. \cite{cat}).

Recent RPA calculations by two independent
groups, using different prescriptions to eliminate the spurious c.m.
strength, give somewhat contradictory results \cite{ham,col}. The
intrinsic isoscalar dipole strength function calculated in
Ref.\cite{col} typically shows two main peaks concentrated around
$10-15~{\rm MeV}$ and $20-30~{\rm MeV}$. The same feature is not
shared by the strength functions calculated in \cite{ham}, where the
low-energy peak has been almost completely eliminated by the
subtraction of the spurious strength. More recent calculations
performed by other groups \cite{shl,gor} give results that are in qualitative
agreement with those of Ref. \cite{col}. However all RPA calculations
start from a  zero-order (single-particle) approximation that does not
treat the
c.m. motion correctly so either perfect self-consistency is required in
order to push all the c.m. strength at zero energy, or some procedure
must be employed to subtract the spurious strength.

A correct description of the c.m. motion is vital when
evaluating the response to the external field (\ref{spur}), since that
operator excites maily the c.m. motion. Here we tackle the problem of
separating the spurious c.m. excitation from the physically
interesting intrinsic excitation by using a semiclassical method based on the
Vlasov equation.
In Ref.\cite{abr},
a solution of the linearized Vlasov equation for finite systems with
moving surface has been derived, which is well suited to describe c.m.
motion (see also Ref.\cite{adm}). Clearly quantum calculations are,
in principle, more rigorous than a semiclassical one, but our hope is that the
physical insight allowed for by the semiclassical method will make our
results more transparent.

We calculate the response of nuclei to an external field of the kind
\beq
\label{vex}
V_{ext}({\bf r},t)=\beta \delta (t) r^{3}Y_{1M}(\bf{\hat r})
\eeq
within the model of Ref.\cite{abr}. In that model the field
(\ref{vex}) does not give spurious dissipation at positive energy
because, like in a perfectly self-consistent RPA calculation, the c.m.
excitation strength is concentrated at zero energy, exactly.
Moreover, in the model of \cite{abr}, contrary to RPA, the c.m. response is
treated correctly already in the starting zero-order approximation.

The RPA (including its relativistic generalizations \cite{vre,pie,pie2}) is
only one of
the two main approaches that have been used to
study the dipole compression mode, the other one being the fluid-dynamic
approach \cite{str,kol}. Our present kinetic-theory approach can be
viewed as intermediate between RPA and fluid dynamics.

\section{Formalism}

In this paper we need to evaluate the following kind of response
functions:
\beq
\label{res}
\tilde{{\cal R}}_{jk}(\omega)=\frac{1}{\beta}
\int d{\bf r}\, r^{j}Y_{1M}({\bf \hat r})\,\delta
\varrho_{k}({\bf r},\omega),
\eeq
more specifically, we need the functions $\tilde{{\cal
R}}_{11},~\tilde{{\cal R}}_{13},~\tilde{{\cal R}}_{33}$ (we put a
tilde over the moving-surface response functions to distinguish them
from the untilded fixed-surface ones).
The fluctuation $\delta \varrho_{k}({\bf r},\omega)$ is the
time Fourier transform of the density fluctuation $\delta
\varrho_{k}({\bf r},t) $ induced by an external field
$V_{ext}({\bf r},t)=\beta \delta(t) r^{k}Y_{1M}(\bf{\hat r})$. This
fluctuation  can be
obtained by integrating over momentum the phase-space density
fluctuation $\delta n_{k}({\bf r},{\bf p},t)$ that is given by the
solution of the linearized Vlasov equation, either with fixed-surface
\cite{bri} or with moving-surface \cite{abr,adm} boundary conditions:
\beq
\delta \varrho_{k} ({\bf r},t)=\int d{\bf p}\,\delta n_{k}({\bf r},{\bf p},t).
\eeq
The solution of the linearized Vlasov equation, both with fixed and
moving surface, can be obtained at different levels of approximation.

Here we consider first a zero-order approximation, obtained by
neglecting the nucleon-nucleon interaction in the bulk. Note that,
while the zero-order approximation of Ref. \cite{bri} corresponds to
nucleons moving in the static equilibrium mean field and being
reflected by the static equilibrium surface, the zero-order
approximation of \cite {abr} and \cite {adm} does take into account
partially the density fluctuations induced by the external driving
force by considering particle reflections from a moving surface.
Thus, what has been called the zero-order approximation in \cite {adm}
includes more dynamics than the corresponding approximation of \cite {bri}.

When working with a moving surface it is convenient to introduce a
modified density fluctuation $\delta \bar{\varrho}$ that is related
to $\delta \varrho$ in the following way. The quantities we are interested
in are
of the kind ($F$ is an arbitrary function of coordinates)
\beq
\delta F(t)
=\int_{V(t)}d{\bf r} F({\bf r}) \varrho ({\bf
r},t) ~-
~\int_{{V_{0}}} d{\bf r} F({\bf r}) \varrho_{0},
\eeq
where $\varrho_{0}$ is the equilibrium density, $V_{0}$ the
equilibrium volume, and $V(t)$ the perturbed time-dependent volume.
Consider the integral
\beq
\int_{V(t)}d{\bf r} F({\bf r}) \varrho ({\bf
r},t) =\int_{V_{0}}d{\bf r} F({\bf r})\varrho_{0}~+~
\int_{\Delta V(t)}d{\bf r} F({\bf r})\varrho_{0}~+~
\int_{V_{0}}d{\bf r} F({\bf r})\delta \varrho({\bf r},t) ~+~{\cal O}{\Big (}
V_{ext}^{2}{\Big)},
\eeq
where $\Delta V(t)=V(t)-V_{0}$; we can include the contribution of
$\int_{\Delta V(t)}d{\bf r} F({\bf r})\varrho_{0}$ into
$\int_{V_{0}}d{\bf r} F({\bf r})\delta \varrho({\bf r},t)$ if we
define \cite{jen}
\beq
\label{rot}
\delta \bar{\varrho} ({\bf r},t)=\delta \varrho({\bf r},t) +\varrho_{0}
\delta (r-R) \delta R({\bf\hat r},t),
\eeq
with $\delta R({\bf \hat r},t)$ giving the fluctuation of the
equilibrium radius $R$ induced by the external force.
Thus, when evaluating the response functions (\ref{res}) we shall
use $\delta \bar{\varrho}$ instead of $\delta\varrho$
 and integrate over the equilibrium volume $V_{0}$ only.

In the zero-order approximation of \cite{adm} the density fluctuation
$\delta\varrho({\bf r},\omega)$ is given by the sum of two terms,
one corresponding to the fixed-surface solution of the Vlasov
equation, and one giving the moving-surface contribution:
\beq
\label{delro}
\delta \tilde \varrho^{0}({\bf r},\omega)=
\delta \bar \varrho^{0}({\bf r},\omega) +
\delta \varrho^{0}_{surf}({\bf r},\omega).
\eeq
The structure of the last term in this equation is similar to that of
the last term in Eq.(\ref{rot}) since $\delta \varrho^{0}_{surf}$ is also
proportional to the fluctuation $\delta R$ of the nuclear surface
(cf. Eq. (3.20) of Ref.\cite{adm}), so it
is convenient to group the two terms together and write
\beq
\label{rze}
\tilde{{\cal R}}_{jk}^{0}(\omega)=
{\cal R}_{jk}^{0}(\omega)+ {\cal S}_{jk}^{0}(\omega)
\eeq
for the response functions in Eq.(\ref{res}).

Here
\beq
{\cal R}_{jk}^{0}(\omega) =
\int dx x^{2} \int dy y^{2} x^{j}y^{k} D^{0}_{L=1}(x,y,\omega)
\eeq
are the zero-order fixed-surface response function calculated according to
Ref.\cite{bri} (the propagator $D^{0}_{L}(x,y,\omega)$ has been
defined in \cite{bri}), while the quantities
\beq
\label{rti}
 {\cal S}_{jk}^{0}(\omega)=\varrho_{0} R^{j+2}\delta
 R^{0}_{1M}(\omega)+\int d{\bf r}\, r^{j}Y_{1M}({\bf \hat
r})\,\delta
\varrho^{0}_{surf}({\bf r},\omega),
\eeq
 include the contributions both of
the last term in Eq.(\ref{rot}) and of the surface term
$\delta \varrho^{0}_{surf}$ in Eq.(\ref{delro}),
the quantity $\delta R^{0}_{1M}(\omega)$ is given in
Eq.(3.21) of Ref.\cite {adm}. According to that equation
\beq
\label{rflu}
\delta R^{0}_{1M}(\omega)=-\beta
\frac{\chi^{0}_{k}(\omega)}{\chi_{1}(\omega)}\,.
\eeq
The functions $\chi^{0}_{k}(\omega)$ and $\chi_{1}(\omega)$ are given
in the Appendix,where the detailed expressions of all the response
functions needed in our calculation are also reported.

\section{C.M. Response}

Before evaluating the intrinsic excitation strength associated with
the external field (\ref{vex}), we prove our previous statement that
the
model of Ref.\cite{abr}, unlike that of Ref.\cite{bri}, gives a
correct description of c.m. motion already at zero order.
From Eq. (\ref{a18}) we have

\beq
\tilde {\cal R}^{0}_{11}(\omega)=\frac{3}{4\pi}\frac{A}{m\omega^{2}}\,,
\eeq
as should be expected for a free particle of mass $Am$ (see for example
\cite{mar}). Since this response function has no poles for $\omega\neq
0$, it does not give spurious dissipation at positive $\omega$.

Another interesting response function is $\tilde {\cal
R}^{0}_{13}(\omega)$, which, in the present moving-surface model is
given by (cf. Eq. (\ref{a21}) )
\beq
\label {r13}
\tilde {\cal R}^{0}_{13}(\omega)=R^{2}~\tilde {\cal R}^{0}_{11}(\omega)\,.
\eeq

The function $\tilde {\cal R}^{0}_{13}(\omega)$ has a direct physical
interpretation as the
displacement of the nuclear c.m. induced by the external field
(\ref{vex}):
\beq
\label {3.3}
\delta R^{(3)}_{c.m.}(\omega)=\frac{\beta}{A}
\tilde {\cal R}^{0}_{13}(\omega)\,.
\eeq
Van Giai and Sagawa \cite{VGS} have proposed a method for removing
the spurious strength associated with the field $r^{3}Y_{1M}({\bf
\hat r})$. They suggested to use an effective external field of the
kind $(r^{3}-\eta r)Y_{1M}({\bf \hat r})$ and to determine the
parameter $\eta$ by the requirement that the external field does not
induce any displacement of the nuclear c.m.:
\beq
\label{eff}
\delta R^{(eff)}_{c.m.}(\omega)=0\,.
\eeq
By using this prescription in a fluid-dynamic model, they determined
$\eta=\frac{5}{3}<r^{2}>$,
where $<r^{2}>$ is the nucleus mean square radius (in our model
$<r^{2}>=\frac{3}{5}R^{2}$). Here we do not
need to use the effective field $(r^{3}-\eta r)$ because the
moving-surface model
employed does not give any spurious dissipation, however, if we do
this, we find
\beq
\delta R^{(eff)}_{c.m.}(\omega)=\frac{\beta}{A}[\tilde {\cal
R}^{0}_{13}(\omega)
-\eta~\tilde {\cal R}^{0}_{11}(\omega)]\,,
\eeq
giving
\beq
\eta=\frac{\tilde {\cal R}^{0}_{13}(\omega)}
{\tilde{\cal R}^{0}_{11}(\omega)}\,.
\eeq
 Using Eq.(\ref{r13}) we get
$\eta=R^{2}$, in agreement with \cite{VGS}.
However we note that this simple result is obtained in a model
that already has the correct translation-invariance property and that
this procedure with $\eta=R^{2}$ cannot be used as a general prescription for
subtracting the spurious strength in non-translationally-invariant
models. Had we used this prescription within the fixed-surface model of Ref
\cite{bri}, that can be taken as an example of a case in which the
problem of spurious dissipation appears, we would have found
\beq
\label{eta}
\eta=\frac{{\cal R}^{0}_{13}(\omega)}
{{\cal R}^{0}_{11}(\omega)}\,,
\eeq
but now the two response functions in the numerator and denominator
are no longer proportional (cf. Eq.(\ref{fsr})), and we would need a
complex and $\omega$-dependent parameter $\eta$ to subtract the spurious
strength.

In order to allow for a more quantitative appreciation of spurious
effects, in Fig.1 we show the zero-order response function
calculated in the fixed-surface approximation of Eq.(\ref{fsr}), that
is with no corrections for c.m. effects (dotted curve), and the same
response function calculated for the effective operator
$(r^{3}-R^{2}r)$ (dashed curve). This effective operator
drastically reduces the response (by almost
90\% for $A=208$), but it leaves the intrinsic strength at the same energy
of the
uncorrected response. The solid curve in Fig.1 instead shows the
result of our present moving-surface model, to be discussed more
extensively in the next section. The quantities plotted in this and in
the two following figures are the strength functions $S(\hbar \omega)$,
related to the respective response functions by
$S(\hbar \omega)=-\frac{1}{\pi}{\rm Im}{\cal R}(\hbar\omega)$. In the
translation-invariant model the strength function includes a
$\delta$-function at the origin,which is not shown. The excitation
energy is $E=\hbar \omega$.

\section{Intrinsic dipole strength}
\subsection{Confined Fermi gas}

After having shown that the model of Ref.\cite{abr} does not introduce
any spurious dissipation at $\omega \neq 0$, we are now in a position
to evaluate the nuclear response to the external field (\ref{vex}),
without having to worry about spurious effects. The response function
$\tilde {\cal R}_{33}^{0}$ is given by Eq.(\ref{a19}). This response function
is qualitatively different from those discussed in the previous section.
Like $\tilde{\cal R}_{11}^{0}$ and $\tilde {\cal R}_{13}^{0}$, it does
contain a part that
is proportional to $\frac{1}{\omega^{2}}$ (given by Eq. (\ref{a20}))
and represents c.m. excitation. This part
does not give any strength at $\omega\neq 0$. In addition to the c.m.
response, $\tilde {\cal R}_{33}^{0}$ contains also an intrinsic part given by
Eq.(\ref{main}). In terms of the variable $\omega$ the intrinsic response
function (\ref{main}) reads
\beq
\label{bella}
\tilde {\cal R}_{intr}^{0}(\omega)={\cal R}_{33}^{0}(\omega)
-\frac{3A}{4\pi}\frac{R^{4}}{m}\frac{1}{\omega^{2}}
{\Big \{}1-{{{\Big [}1-\frac{1}{2}\omega^2
\frac{{\cal R}_{13}^{0}(\omega)}{m_{13}^{1}}
{\Big]}^2}\over{1-\frac{1}{2}\omega^2
\frac{{\cal R}_{11}^{0}(\omega)}{m_{11}^{1}}}}
{\Big\}}\,,
\eeq
with the $\omega$-moments $m_{jk}^{p}$ related to the $s$-moments
(\ref{mom}) by
\beq
m_{jk}^{p}={\big (}\frac{v_{F}}{R}{\big )}^{p+1}{\cal M}_{jk}^{p}\,.
\eeq
Equation (\ref{bella}) does
express the intrinsic response function associated with the
operator (\ref{spur}) in terms of response functions calculated in the
underlying
non-translationally-invariant model, the analogous expression given
by the widely used prescription $r^{3}\to (r^{3}-\eta r)$ (with real $\eta$) is
\beq
\label{reff}
{\cal R}_{eff}^{0}(\omega)={\cal R}_{33}^{0}(\omega)-
2\eta {\cal R}_{13}^{0}(\omega)+
\eta^{2}{\cal R}_{11}^{0}(\omega).
\eeq
We can easily check
that, if we were allowed to make the approximation ${\cal
R}^{0}_{13}(\omega)=
 R^{2}{\cal R}^{0}_{11}(\omega)$, which is
not valid in the fixed-surface model, then Eq. (\ref{bella}) would
be equivalent to Eq. (\ref{reff}), but the two formulae are
different in general.The intrinsic strength function associated
with the response function (\ref{bella}) is shown by the solid curves
in Figs. 1 and 2, while that given by Eq.(\ref{reff}) with
$\eta=R^{2}$ is given by the dashed curves in the same figures.
The strength distributions given by the two formulae (\ref{bella})
and (\ref{reff})are qualitatively different, in spite of the fact
that they give the same moments $m^{1}$ and $m^{-1}$.
Actually, using the high-frequency expansion \cite{lip}
\beq
{\cal
R}_{jk}^{0}(\omega)|_{\omega\to\infty}=\frac{2m^{1}_{jk}}{\omega^{2}}+{\cal
O}
(\frac{1}{\omega^{3}})\,,
\eeq
Eq. (\ref{bella}) gives
\beq
\label{ews}
\int_{0}^{\infty}d(\hbar\omega) \hbar\omega{\Big [}-\frac{1}{\pi}
{\rm Im}\tilde {\cal R}_{intr}^{0}(\omega){\Big ]}=\frac{3\hbar^{2}}
{14\pi m}A R^{4}
\eeq
for the intrinsic energy-weighted sum rule $\tilde m^{1}_{intr}$ associated
with the
operator (\ref{spur}), in agreement with Eq. (A.4) of Ref. \cite{VGS}
(in our model $<r^{4}>=\frac{3}{7} R^{4}$).

The $m^{-1}$ moment is also the same for the two response functions
(\ref{bella}) and (\ref{reff}). Taking the
limit of small frequency, we get
\beq
\label{hydro}
\lim_{\omega\to 0} \tilde {\cal R}_{intr}^{0}(\omega)=
\lim_{\omega\to 0}  {\cal R}_{eff}^{0}(\omega)=
-\frac{6}{35\pi}\frac{AR^{6}}{K_{FG}}\,,
\eeq
where $K_{FG}=6\epsilon_{F}$ is the Fermi-gas incompressibility
parameter.
 This limit is related to the hydrodynamic sum rule
\beq
\tilde m_{intr}^{-1}=\int_{0}^{\infty}d
\omega \frac{1}{ \omega}{\big [}-\frac{1}{\pi}{\rm
Im}\tilde {\cal R}_{intr}^{0}(\omega){\big ]}
\eeq
by \cite{lip}
\beq
\lim_{\omega\to 0} \tilde {\cal R}_{intr}^{0}(\omega)=-2\tilde
m_{intr}^{-1}\,.
\eeq
Having established the value of the two moments $\tilde
m_{intr}^{1}$ and $\tilde m_{intr}^{-1}$, we can calculate the mean
excitation energy as
\bea
\hbar \tilde \omega_{1-1}&=&\sqrt{\frac{\tilde m_{intr}^{1}}{\tilde
m_{intr}^{-1}}}\nonumber \\
&=&\hbar \sqrt{\frac{3}{2}\frac{K_{FG}}{m<r^{2}>}}\,,
\eea
in agreement with \cite{str}, where a phenomenological
incompressibility parameter $K$ was used instead of $K_{FG}$. Since
$K_{FG}\approx 200~{\rm MeV}$ (for $R=1.2~A^{\frac{1}{3}}{\rm fm}$), while the
phenomenological value
used in \cite{str} was $K=220 ~{\rm MeV}$, we see that, already at the
Fermi-gas level, we have a resonable estimate for the mean excitation energy
of the isoscalar dipole strength. What is unrealistic in the Fermi gas
model, is the pressure $P$ exerted by the nuclear fluid, but the
incompressibility, defined as $K\equiv 9\frac{\partial P}{\partial
\varrho}$, happens to be within the range of $210\pm 30~ {\rm MeV}$
compatible with the monopole data \cite{bla}. In any case we can always
change the compressibility of the system by switching on the residual
interaction between nucleons.

\subsection{Changing the compressibility}

So far we have treated the nucleus like a gas of non-interacting
fermions confined to a spherical cavity with perfectly reflecting
walls that are allowed to move under the action of the gas pressure
induced by the external field,
taking into account the interaction between nucleons changes the
compressibility of this nuclear fluid.

Rather than embarking in numerical calculations, here we give only a
simple estimate of the main effects that are expected when the
nucleon-nucleon interaction is switched on and, in order to keep the
analytic insight allowed for by our semiclassical approach, we assume
a very schematic, separable, effective interaction of the
dipole-dipole type:
\beq
\label{dip}
u({\bf r},{\bf r}')=\alpha \sum_{M}r r'\,Y_{1M}(\hat{\bf r})
Y_{1M}^{*}(\hat{\bf r}').
\eeq

The parameter ${\cal F}_{0}$, determining the strength of the interaction,
will be related to the compressibility, which is the relevant
parameter in this context. Then the fixed-surface collective
response functions${\cal R}_{1k}(\omega)$,
given by the solution of the RPA-type integral equation (5.28) of
Ref. \cite{bri} are:
\beq
\label{finfs2}
{\cal R}_{1k}(\omega )=
\frac{{\cal R}_{1k}^{0}(\omega )}
{1- \alpha {\cal R}_{11}^{0}(\omega )}
~~~~~~~~~~~~~~~~~~~(k=1,3)\,,
\eeq
and
\beq
\label{finfs}
{\cal R}_{33}(\omega )=
{\cal R}_{33}^{0}(\omega )
+\alpha \frac{{\big [}{\cal R}_{13}^{0}(\omega ){\big ]}^2}
{1- \alpha {\cal R}_{11}^{0}(\omega )}.
\eeq
Now, if we replace the Fermi gas fixed-surface response functions
${\cal R}_{jk}^{0}(\omega )$ with the above
collective response functions ${\cal R}_{jk}(\omega)$
in Eq.(\ref{bella}), we obtain a new intrinsic response function
\bea
\label{collintr}
\tilde {\cal R}_{intr}(\omega)&=&
{\cal R}_{33}^{0}(\omega )
+\alpha  \frac{{\big [}{\cal R}_{13}^{0}(\omega ){\big ]}^2}
{1- \alpha {\cal R}_{11}^{0}(\omega )} \nonumber \\
&-&\frac{3A}{4\pi}\frac{R^{4}}{m}\frac{1}{\omega^{2}}
{\Big \{}1-{{{\Big [}1-\frac{1}{2}\omega^2
\frac{{\cal R}_{13}^{0}(\omega)}{m_{13}^{1}}
\frac{1}{1-\alpha {\cal R}_{11}^{0}(\omega)}
{\Big]}^2}\over{1-\frac{1}{2}\omega^2
\frac{{\cal R}_{11}^{0}(\omega)}{m_{11}^{1}}
\frac{1}{1-\alpha{\cal R}_{11}^{0}(\omega)}}}
{\Big\}}\,.
\eea
The energy-weighted moment $\tilde m^{1}_{intr}$ of this response
function is still given by Eq.(\ref{ews}), as can be easily checked by
studying the large-$\omega$ limit. Thus we are confident that, even if
the dipole-dipole residual interaction (\ref{dip}) violates translation
invariance, we are not introducing any spurious strength in our
intrinsic response function (\ref{collintr}).

Contrary to the first moment $m^{1}$, the inverse moment $m^{-1}$ is
altered by the residual interaction. This can be checked by evaluating
the limit
\beq
\label{m-1}
\lim_{\omega \to 0} \tilde{\cal R}_{intr}(\omega)= -\frac{6}{35
\pi}\frac{AR^{6}}{6\epsilon_{F}[1-\alpha {\cal
R}_{11}^{0}(0)]}[1-\frac{5}{14}\alpha {\cal R}_{11}^{0}(0)]\,,
\eeq
where
\beq
\label{led}
{\cal R}_{11}^{0}(0)=-\frac{2}{5}\frac{9A}{16\pi}\frac{R^{2}}{\epsilon_{F}}\,.
\eeq
Equation (\ref{m-1}) can be written as
\beq
\lim_{\omega \to 0} \tilde{\cal R}_{intr}(\omega)= -\frac{6}{35
\pi}\frac{AR^{6}}{K}\,,
\eeq
with
\beq
K=6\epsilon_{F}\frac{1-\alpha {\cal
R}_{11}^{0}(0)}{1-\frac{5}{14}\alpha {\cal R}_{11}^{0}(0)}\,.
\eeq

By following exactly the same procedure for monopole vibrations, for
which we assume a monopole residual interaction of the kind
\beq
u(r,r')=\frac{\beta}{4\pi}r^{2}~r'^{2}\,,
\eeq
we find an analogous expression for the monopole incompressibility:
\beq
\label{mcomp}
K_{mon}=6\epsilon_{F}\frac{1-\beta{\cal
R}_{L=0}^{0}(0)}{1-\frac{21}{26}\beta{\cal R}_{L=0}^{0}(0)}\,,
\eeq
where ${\cal R}_{L=0}^{0}(0)$ is the zero-frequency limit of the
monopole response function, analogous to (\ref{led}).
We can determine $K_{mon}$ by comparison with the data on giant
monopole resonances (see Table I) and then, by assuming $K=K_{mon}$,
we can study the dipole response.

 In our simplified scheme for taking into account the residual
interaction between nucleons, we replace the Fermi-gas intrinsic
response function (\ref{bella}) with Eq. (\ref{collintr}). Thus
Eq.(\ref{collintr})
becomes our final expression for the intrinsic isoscalar dipole
response function, it is a generalization of the confined-Fermi-gas
expression (\ref{bella}) to a confined fluid of interacting nucleons
with incompressibility $K$. At first we treat $K$ as a free parameter
and study its effect on the dipole response function, then we
determine the value of $K$ by comparison with the monopole-resonance
data and use this value to compare with the dipole-resonance data.

 The strength function associated with the
response function (\ref{collintr}) is shown in Fig. 3 for three
values of the incompressibility ($K=K_{FG}$, $K<K_{FG}$, and
$K>K_{FG}$).
When the system becomes softer, part of the strength
shifts from the high-energy peak towards the low-energy peak, at about
13 MeV in Lead. The position of the two centroids is also
slightly changed, but the main effect is the change in the relative
weight of the two peaks.This behaviuor is similar to that found in
the relativistic approach of Ref. \cite{vre}, the position of the
peaks is also similar for comparable values of $K$. However, unlike
Ref. \cite{vre}, we do not see any qualitative difference in the
sensitivity of the two peaks to the compressibility of the system.

We finally note that for $K>K_{FG}$ the strength moves to
higher energy.

\subsection{Comparison with data}

In Table I we report experimental centroid energies of the giant
monopole resonance taken from Ref. \cite{you}, together with our
calculated values. A reasonable agreement with experiment is obtained
for $K_{mon}=180$ MeV.

In Fig.4 we compare our enegy-weighted strength functions
with the recent data of Ref.\cite{cla} for $^{208}{\rm Pb}$,
$^{116}{\rm Sn}$
and $^{90}{\rm Zr}$.
The quantity plotted in the figure is
\beq
y(E)=100E {\Big (}-\frac{1}{\pi}{\rm Im}
\tilde {\cal R}_{intr}(E){\Big )}/\tilde m^{1}_{intr}\,,
\eeq
where $E=\hbar\omega$ is the excitation energy.
The solid curve shows our confined-Fermi-gas
response, that is the energy-weighted isoscalar dipole strength for an assembly
of non-interacting
nucleons with incompressibility $K\approx 200 ~{\rm MeV}$. Our strength
integrated up to $\hbar\omega =40 ~{\rm MeV}$ exhausts about $90\%$ of
the energy-weighted sum rule $\tilde m^{1}_{intr}$, while integrating
up to $100 ~{\rm MeV}$ exhausts about $99\%$ of the energy-weighted sum
rule. It is important to note that the data
overshoot this sum rule \cite{cla}. The reasons for this overshooting
are not clear to us, since, unlike the isovector dipole, no
enhancement factor of the energy-weighted sum rule is expected in
this case \cite{lip}.

The experimental finding of \cite{cla} that the isoscalar dipole strength
consists of two components is qualitatively reproduced by our
semiclassical model  ( at the Fermi gas level the two resonances correspond to
two complex zeroes of the function $\chi_{1}(s)$ of
Eq.(\ref{closed})\cite{adm}). In this respect our results are also in
qualitative agreement with the quantal RPA calculations of
\cite{col,shl,gor} and \cite{vre,pie,pie2},that predict a two-resonance
structure in the isoscalar dipole response.

 For the confined Fermi gas (solid line) the
position of the two peaks is too high in energy.
Taking into account the residual interaction (dashed line)
improves the agreement with data, but does not completely
remove the discrepancy, especially for the high-energy peak.

The dashed curve in Fig.4 has been calculated for
 $K=180$ MeV, which has been
determined by fitting the energy of the isoscalar monopole resonance
within the same model.

Our value of $K$ lies at the lower end of the uncertainty range
indicated in Ref. \cite{bla} and is somewhat smaller than the values
reported in Ref. \cite{ber}. However these values refer to saturated and
symmetric ($N=Z$) nuclear matter. The small discrepancy is
probably due to the fact that we are using a sharp-surface model,
while of course experiments are performed on real nuclei that have a
diffuse surface, thus our value of $K$ does not refer to the density
in the interior of nuclei, but rather to some average density that
takes into account surface diffuseness. Also, in real nuclei $N$ is larger
than $Z$ and,
as pointed out in \cite{yos}, $K$ decreases with increasing asymmetry.

By comparing the two curves shown in Fig.4, we can say that an attractive
interaction between nucleons, that decreases the value of
$K$ with respect to the initial value of $200 {~\rm MeV}$,  improves the
agreement with data, but the position of the high-energy peak remains
too high in energy, even for the rather small value $K=180 ~{\rm MeV}$,
suggested by our fit to monopole data. Further decrease of $K$ would
shift the response towards lower energy and would
redistribute the strength by enhancing the low-energy peak and
depleting the high-energy one, moreover this would give too small
values for the monopole-resonance centroid energies.
The coupling of the two resonances in the isoscalar dipole response due to the
residual interaction is a novel effect displayed by our
calculation and, although the position of the  peaks is not
much affected by the compressibility, we find that the stregth
associated with them is quite sensitive to the value of $K$.

Another qualitative feature of the data
that is well reproduced by our calculations is the $A^{-\frac{1}{3}}$
dependence of the peak positions \cite{cla}. Actually, an interesting
property of
the response function (\ref{collintr}) is that, if $K/\epsilon_{F}$ is
independent of $A$, then this quantity, like the functions
${\cal R}_{jk}^{0}$, can be expressed in terms of a universal
function of the dimensionless parameter $s$ defined in the Appendix:
\beq
\label{sca}
\tilde {\cal R}_{intr.}(s)={\cal N}_{33}(A) u(s).
\eeq
The function $u(s)$ is the same
for all spherical nuclei and its explicit expression can be easily
derived from the equations given in the Appendix.
The mass number $A$ of the particular nucleus being studied enters
only through the normalization factor
\beq
{\cal N}_{33}(A)=\frac{9A}{16 \pi}\frac{R^{6}}{\epsilon_{F}}
\eeq
and does not affect the position of the resonances in the variable $s$.
Since $s\propto A^{\frac{1}{3}}$, the scaling property (\ref{sca}) of the
response
function (\ref{collintr}) results in an $A^{-\frac{1}{3}}$ dependence of
the peak positions in excitation energy.

In conclusion, our model is able to reproduce some of the
features displayed by the data like the two-resonance structure of
the response and the $A$-dependence of the peak position, but, even after
an accurate subtraction
of the spurious c.m. strength for which a new formula has been derived,
it fails to reconcile the monopole and dipole data. This failure is shared
by our
semiclassical model with the quantal RPA calculations of Ref.s
\cite{ham,col,shl,gor} and \cite{vre,pie,pie2}.

\section{Acknowledgements}

V.I.A. has been partially supported by INFN and MURST, Italy. A.D. is grateful
for the warm hospitality extended to him during his visit to
the Institute for Nuclear Research, Kiev.

\newpage

\appendix
\section*{Response functions}

In this Appendix we give explicit expressions for the
functions $\tilde {\cal R}^{0}_{11},~\tilde {\cal
R}^{0}_{13}$ and $\tilde {\cal R}^{0}_{33}$. Instead of the variable
$\omega$, it is convenient to use the dimensionless variable
\beq
s=\frac{\omega}{v_{F}/R}\,,
\eeq
where $v_{F}$ is the Fermi velocity, we shall use also the Fermi
energy $\epsilon_{F}$ to characterise our systems.
Moreover we add
a vanishingly small positive imaginary part to the variable $s$ in
order to define the behaviour of the response functions at poles: $s\to
s+i\varepsilon$.

The fixed-surface part of the response functions (\ref{rze}) can be
calculated by using the propagator $D^{0}_{L=1}(r,r',\omega)$ given in
Ref.\cite{bri}. The explicit expression for $A$ non-interacting nucleons
at zero temperature contained in a spherical cavity of radius $R$ is:
\beq
\label{fsr}
{\cal R}_{jk}^{0}(s)=
\frac{9A}{16\pi}\frac{1}{\epsilon_{F}}\sum_{n=-\infty}^{+\infty}
\sum_{N=\pm 1} \int_{0}^{1} dx x^{2}
~s_{nN}(x){{Q^{j~*}_{nN}(x){Q^{k}_{nN}(x)}\over {s-s_{nN}(x)}}}.
\eeq
We have defined
\beq
s_{nN}(x)=\frac{n\pi+N\alpha(x)}{x},
\eeq
and $\alpha(x)=\arcsin(x)$,
while the Fourier coefficients $Q^{k}_{nN}(x)$ are defined according to
Ref.\cite{bri}, they  are the classical limit of the radial matrix elements
of the
operator $r^{k}$. The coefficients needed in our calculations are:
\beq
\label{fou1}
Q^{1}_{nN}(x)=(-)^{n}R
\frac{1}{s^{2}_{nN}(x)}~~~~~~~~~~~~~~~~~~~~~~~~~~~~~~~~~~~~~~~~~~
\eeq
and
\beq
\label{fou3}
Q^{3}_{nN}(x)=(-)^{n}R^{3}\frac{3}{s^{2}_{nN}(x)}{\Big(}1+\frac{4}{3}N
\frac{\sqrt{1-x^{2}}}{s_{nN}(x)}
-\frac{2}{s^{2}_{nN}(x)}{\Big)}\,.
\eeq

The functions $\chi^{0}_{k}(s)$ and $\chi_{1}(s)$, associated with
the moving-surface part of the response (cf. Eq. (\ref{rflu})), are given by
\beq
\label{chik}
\chi^{0}_{k}(s) =\frac{9A}{8\pi}\frac{1}{R^{3}}\sum_{n=-\infty}^{+\infty}
\sum_{N=\pm 1}\int_{0}^{1} dx x^{2}~
s_{nN}(x){(-)^{n}{Q^{k}_{nN}(x)}
\over {s-s_{nN}(x)}},
\eeq
and
\beq
\label{chi1}
\chi_{1}(s)=-\frac{9A}{4\pi}~\epsilon_{F}~s
\sum_{n=-\infty}^{+\infty}
\sum_{N=\pm 1}\int_{0}^{1} dx x^{2}
~{1\over {s-s_{nN}(x)}}.
\eeq

 The function $\chi_{1}(s)$
is the $L=1$ component of  $\chi_{L}(s)$, defined in Eq.(3.23) of
\cite{adm}, we have used the pole expansion of the cotangent to write
it in the present form, while the fuctions $\chi^{0}_{k}(s)$ give
the numerator of Eq. (3.21) of the same reference for
external fields $r^{k}Y_{1M}(\hat{\bf r})$.
We notice the similarity between the fixed-surface
response function (\ref{fsr}) and the two functions $\chi^{0}_{k}(s)$ and
$\chi_{1}(s)$ appearing in the moving-surface part of the response,
in particular the integrand in all these functions has the same
poles, while the numerators are different.

By using the explicit expressions of the Fourier coefficients
(\ref{fou1}) and (\ref{fou3}) and, repeatedly, the identity
\beq
\label{iden}
\frac{1}{s_{nN}}~\frac{1}{s-s_{nN}}=\frac{1}{s}(\frac{1}{s_{nN}}
+\frac{1}{s-s_{nN}})
\eeq
it is possible to express the fixed-surface response functions in
terms of the functions $\chi^{0}_{k}(s)$ and $\chi_{1}(s)$. A useful
identity, obtained in this way, is
\beq
\label{a12}
\frac{2\epsilon_{F}}{R^{4}}{\cal R}_{1k}^{0}(s)=
\frac{1}{s^{2}}{\Big
(}\chi^{0}_{k}(s)-\chi^{0}_{k}(0){\Big)}
~~~~~~~~~~~~~~~~~~~~~~~~(k=1,3)\,.~~~~~~~~~~~~~~~~~~
\eeq
We do not report here the analogous expression for
${\cal R}_{33}^{0}(s)$ since it is too
involved. Moreover
\beq
\label{chizuno}
\chi^{0}_{1}(s)=-\frac{1}{2\epsilon_{F}R^{2}}\frac{\chi_{1}(s)}{s^{2}}\,,
~~~~~~~~~~~~~~~~~~~~~~~~~~~~~~~~~~~~~~~~~~~~~~~~~~~~~~~~~~~~~~~~~~~
\eeq
\beq
\chi^0_3(s) =
-\frac{3}{2\epsilon_{F}}\frac{\chi_{1}(s)}{s^{2}}(1-\frac{2}{s^2})
-\frac{3A}{\pi}\frac{1}{s^{2}}{\Big [}1
-\frac{3}{4}\int_{0}^{1} dx x^2 \sum_{N=\pm 1}
{{\sin(sx+N\alpha(x))}\over{\sin(sx-N\alpha(x))}}{\Big]}\,,
\eeq
and
\beq
\label{chizero}
\chi^{0}_{3}(0)=R^{2} \chi^{0}_{1}(0)=-\frac{3A}{4\pi}\,.
\eeq

The advantage of expressing the fixed-surface response functions and
the functions $\chi^{0}_{k}(s)$ in terms of the function $\chi_{1}(s)$
is due to the fact that the infinite sum over $n$ in $\chi_{1}(s)$
can be performed exactly, consequently we can obtain exact closed expressions
for all these functions within this model. A useful closed expression
for $\chi_{1}(s)$ is
\beq
\label{closed}
\chi_{1}(s)=-3\epsilon_{F}\frac{3A}{4\pi}~s
\int_{0}^{1} dx x^{3}\frac{\sin(2sx)}{\sin^{2}(sx)-x^{2}}\,.
\eeq

Thus, by using the relations given above, we can connect the zero-order
fixed-surface  and moving-surface solutions of the Vlasov equation
obtained in Refs.\cite{abr,bri}.
This is particularly simple for $ {\cal S}_{11}^{0}$ and
${\cal S}_{13}^{0}$ since
\beq
\label{a7}
{\cal S}_{1k}^{0}(s)=
-\frac{R^{4}}{2\epsilon_{F}}\frac{\chi^{0}_{k}(s)}{s^{2}}
~~~~~~~~~~~~~~~~~~~(k=1,3)\,,
\eeq
while it is somewhat more complicated for ${\cal S}_{33}^{0}$
since
\beq
\label{rtil}
{\cal S}_{33}^{0}(s)=
-\frac{R^{4}}{2\epsilon_{F}}\frac{\chi^{0}_{3}(s)}{s^{2}}
\frac{\chi^{0}_{3}(s)}{\chi_{1}(s)}\,.
\eeq
By inserting $\chi^0_k(s)$ obtained from Eq. (\ref{a12}) into (\ref{a7}),
we get
\beq
\label{a17}
{\cal S}_{1k}^{0}(s)=-{\cal R}_{1k}^{0}(s)+
\frac{3A}{4\pi}\frac{R^{k+1}}{2\epsilon_{F}}\frac{1}{s^{2}}
~~~~~~~~~~~~~~~(k=1,3)\,,
\eeq
that gives
\beq
\label{a18}
\tilde {\cal
R}_{11}^{0}(s)=\frac{3A}{4\pi}\frac{R^{2}}{2\epsilon_{F}}\frac{1}{s^{2}}
\eeq
and
\beq
\label {a21}
\tilde {\cal R}_{13}^{0}(s)=R^{2}\tilde {\cal R}_{11}^{0}(s)
\eeq
for two of the full response functions in (\ref{rze}).

For $\tilde {\cal R}_{33}^{0}$, we obtain
\beq
\label {a19}
\tilde {\cal R}_{33}^{0}(s)=\tilde {\cal R}_{c.m.}^{0}(s)+
\tilde {\cal R}_{intr}^{0}(s)\,,
\eeq
with
\beq
\label{a20}
\tilde {\cal
R}_{c.m.}^{0}(s)=\frac{3A}{4\pi}\frac{R^{6}}{2\epsilon_{F}}\frac{1}{s^{2}}
\eeq
and
\beq
\label{main}
\tilde {\cal R}_{intr}^{0}(s)={\cal R}_{33}^{0}(s)
-\frac{3A}{4\pi}\frac{R^{6}}{2\epsilon_{F}}\frac{1}{s^{2}}
{\Big \{}1-{{{\Big [}1-\frac{1}{2}s^2
\frac{{\cal R}_{13}^{0}(s)}{{\cal M}_{13}^{1}}
{\Big]}^2}\over{1-\frac{1}{2}s^2
\frac{{\cal R}_{11}^{0}(s)}{{\cal M}_{11}^{1}}}}
{\Big\}}
\eeq
giving the c.m. and intrinsic response, respectively.
The last equation expresses the intrinsic response function for the
operator (\ref{spur}) in terms of non-translationally-invariant response
functions and is the main result of the present paper.
It is an exact relation within the present model (confined Fermi gas).
The moments
\beq
\label{mom}
{\cal M}_{jk}^{p}=\int_{0}^{\infty}dss^{p}{\big [}-\frac{1}{\pi}{\rm
Im}{\cal R}_{jk}^{0}(s){\big ]}
\eeq
are defined in terms of the fixed-surface response functions and they
can be easily evaluated from (\ref{fsr}). Explicitly:
${\cal M}_{11}^{1}=\frac{1}{3}\frac{9A}{16\pi}\frac{R^{2}}{\epsilon_{F}}$,
${\cal M}_{13}^{1}=R^{2}{\cal M}_{11}^{1}$, and
${\cal M}_{33}^{1}=\frac{11}{7}R^{4}{\cal M}_{11}^{1}$.

An essential property of the intrinsic response function (\ref{main}) is
that its limit for $s\to 0$ is finite.
\newpage


\newpage
\begin{table}
\caption{Experimental and calculated energies of giant monopole
resonance. Data from Ref. [22]}
\label{t:1}
\begin{tabular}{ccccccccc}
Nucleus       & Exp. energy    &  Calc. energy         \\

& MeV &  MeV  	      \\
\tableline
$^{208}{\rm Pb}$    &14.17 $\pm$0.28      &14.1         \\
$^{116}{\rm Sn}$    &16.07 $\pm$0.12      &17.2         \\
$^{90}{\rm Zr}$     &17.89 $\pm$0.20      &18.7         \\
\end{tabular}
\end{table}
\newpage
\centerline{FIGURE CAPTIONS}
\vskip 1.cm
Fig.1~~Isoscalar dipole response in fixed- and moving-surface models.
The dotted line shows  the fixed-surface response for radial dependence
of operator $Q(r)=r^{3}$, while the dashed line is for the effective operator
$Q(r)=r^{3}-R^{2}r$. The solid line shows the zero-order moving-surface
strength given by Eq.(\ref{bella}).
\vskip 1. cm
Fig.2~~ The dashed and solid curves are the same as in Fig.1 . Note
the change of vertical scale.
\vskip 1. cm
Fig.3~~ Dipole strength for different value of incompressibility
parameter. The solid curve shows our Fermi-gas result.
\vskip 1. cm
Fig.4~~ Comparison of our energy-weighted strength
functions with experimental data of Ref. \cite{cla} for $^{208}{\rm
Pb}$ (a), $^{116}{\rm Sn}$ (b) and $^{90}{\rm Zr}$ (c). The solid curve
shows the Fermi-gas result, corresponding to incompressibility
$K\approx 200 ~{\rm MeV}$, the dashed curve is for interacting nucleons
with $K=180 ~{\rm MeV}$.
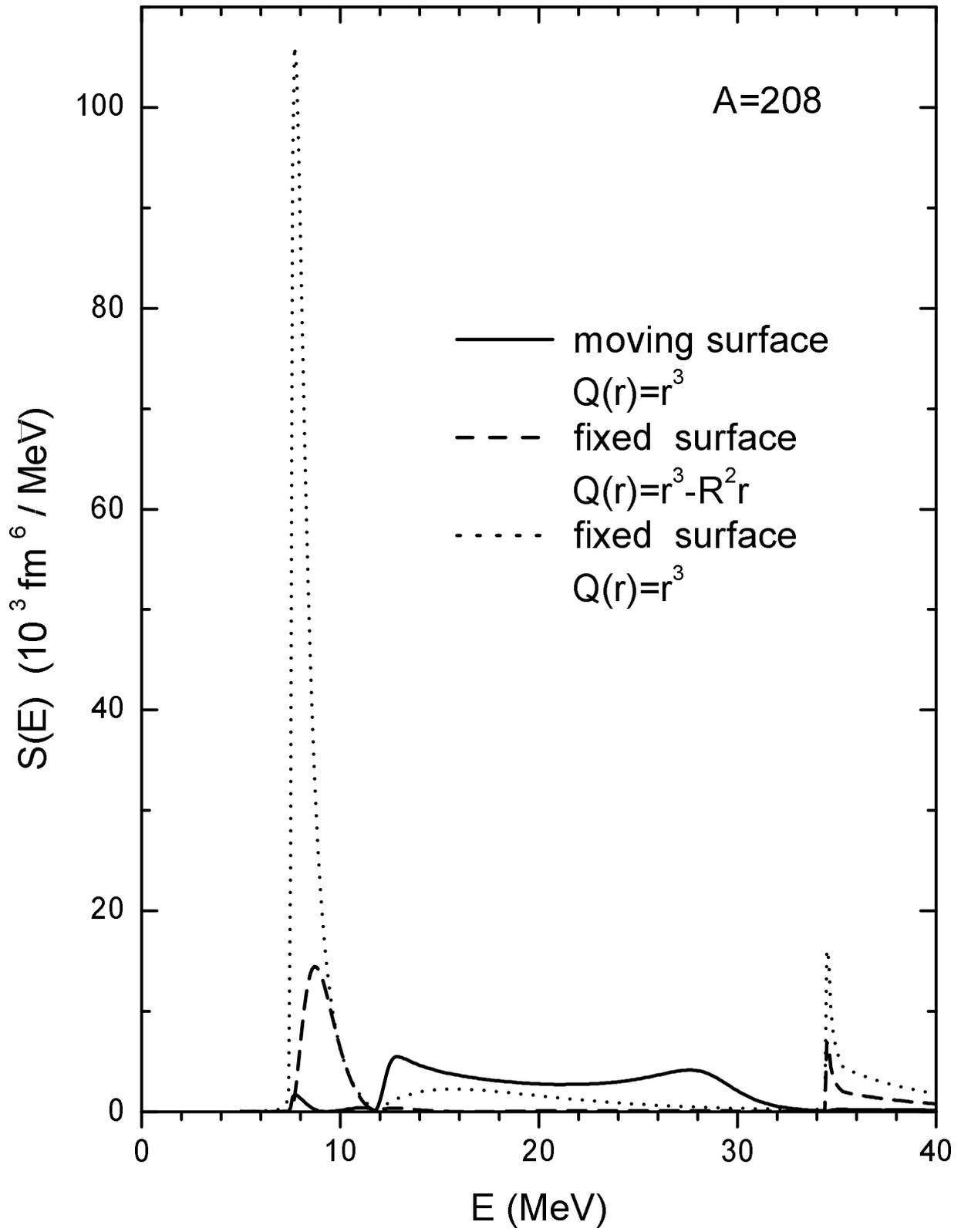
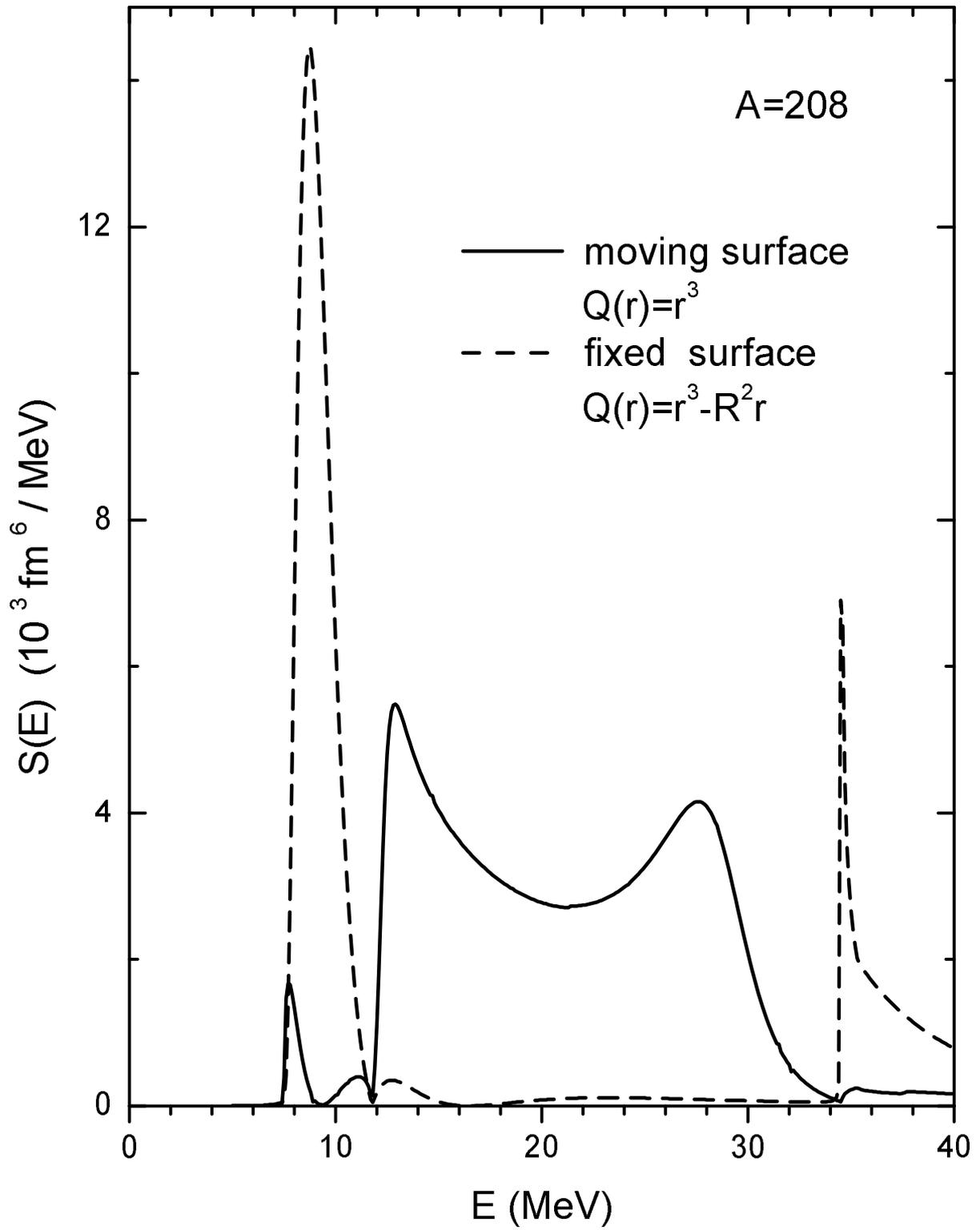
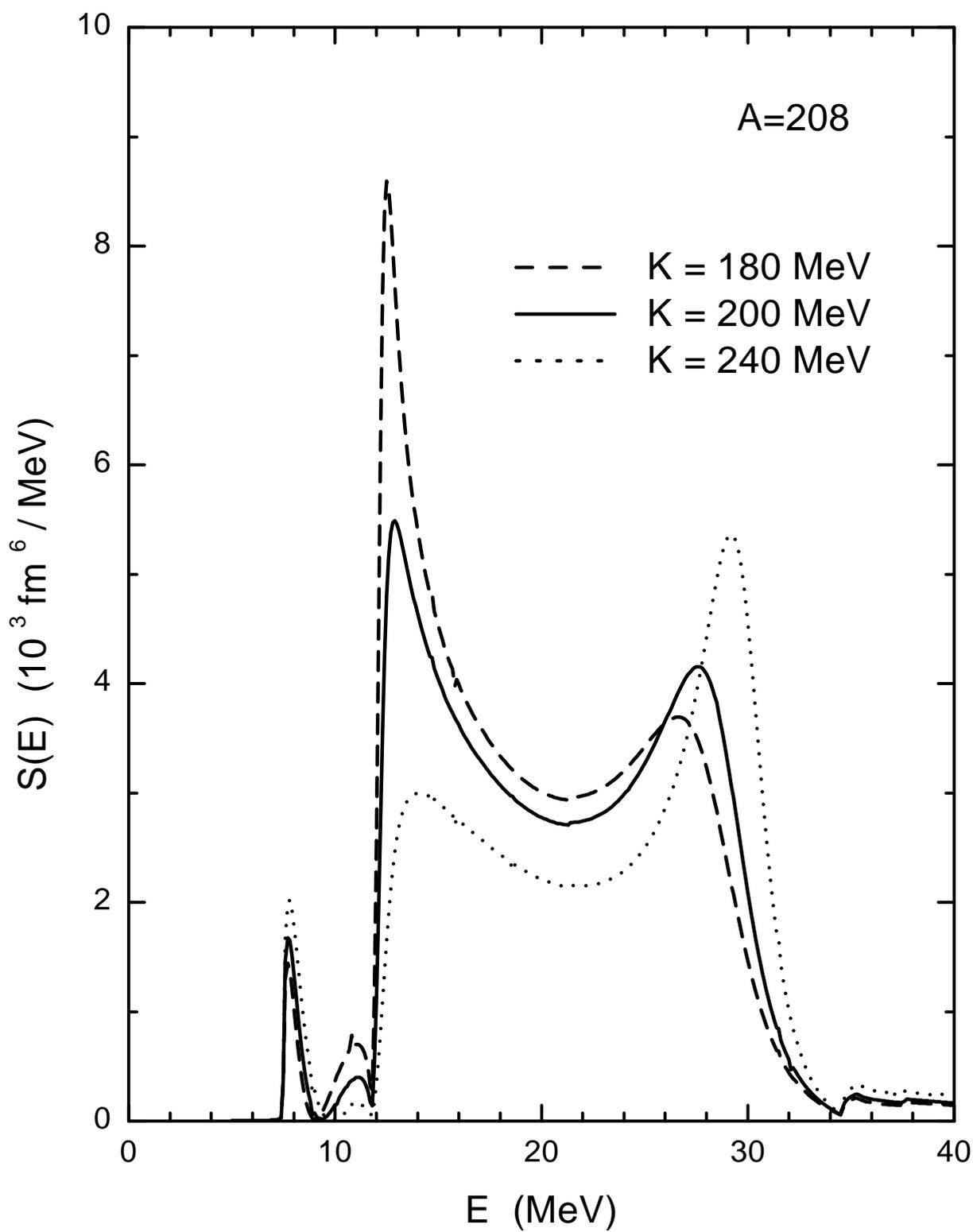
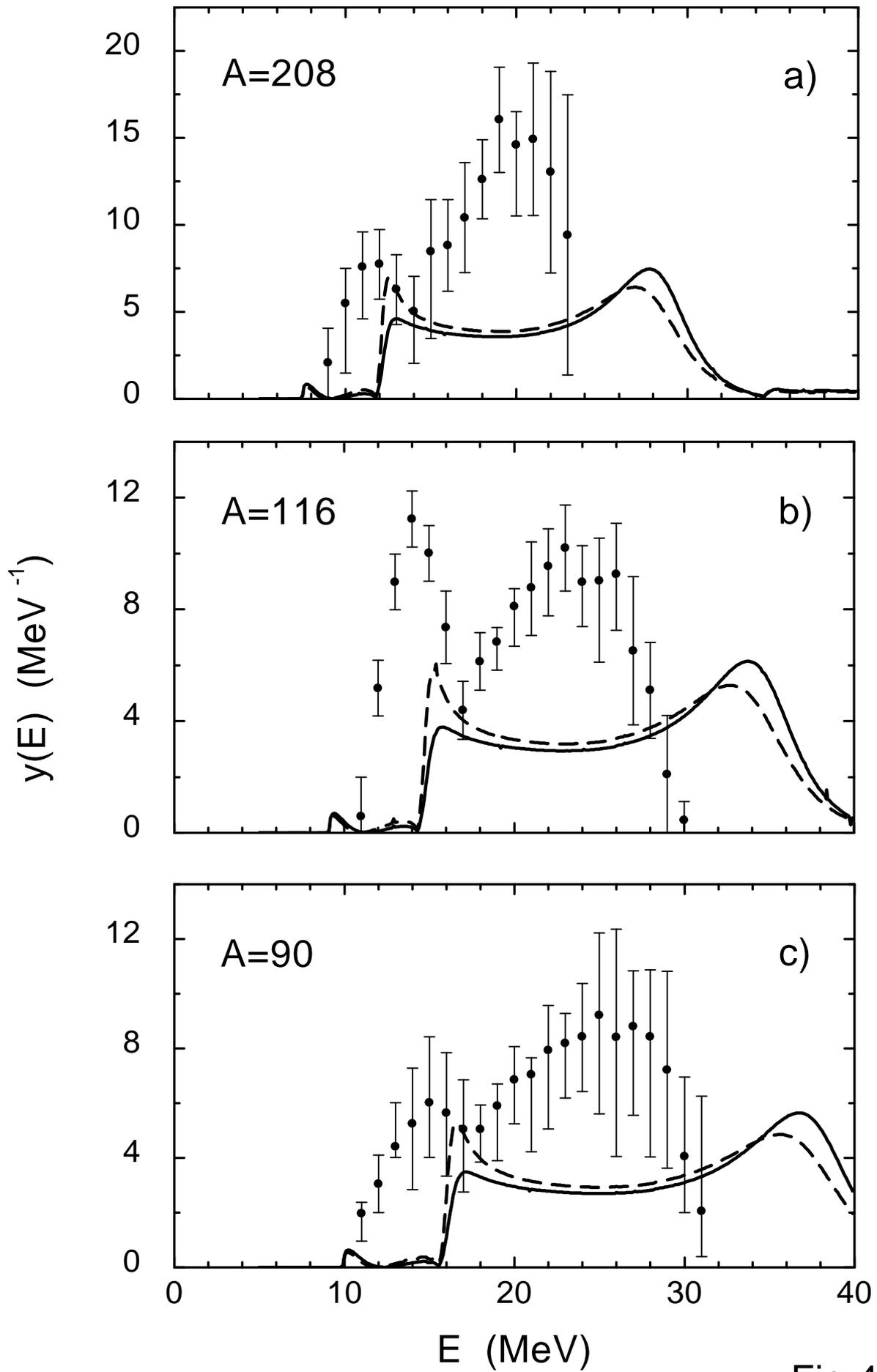
\end{document}